 \documentclass[12pt,preprint]{aastex}

\newcommand{\begeq}{\begin{equation}}
\newcommand{\fineq}{\end{equation}}

\newcommand{\rs}{r_{_{\rm S}}}

\newcommand{\asym}{\doteq}

\newcommand{\OmegaK}{\Omega_{\rm K}}

\newcommand{\gapprox}{\lower.4ex\hbox{$\;\buildrel >\over{\scriptstyle\sim}\;$}}
\newcommand{\lapprox}{\lower.4ex\hbox{$\;\buildrel <\over{\scriptstyle\sim}\;$}}

\def\vr{v}
\def\ellprime0{\ell'_0}

\slugcomment{accepted by \apj}

\begin{document}

\shorttitle{Viscosity Restrictions in ADAF Disks}
\shortauthors{Becker \& Subramanian}

\title{RESTRICTIONS ON THE PHYSICAL PRESCRIPTION FOR THE
VISCOSITY IN ADVECTION-DOMINATED ACCRETION DISKS}

\author{Peter A. Becker\altaffilmark{1}$^,$\altaffilmark{2}}

\affil{Center for Earth Observing and Space Research, \break
George Mason University \break
Fairfax, VA 22030-4444, USA}

\and

\author{Prasad Subramanian\altaffilmark{3}}
\affil{Inter-University Center for Astronomy and Astrophysics \break
P.O. Bag 4, Ganeshkhind, Pune - 411 007, India}

\vfil

\altaffiltext{1}{pbecker@gmu.edu}
\altaffiltext{2}{also Department of Physics and Astronomy,
George Mason University, Fairfax, VA 22030-4444, USA}
\altaffiltext{3}{psubrama@iucaa.ernet.in}

\date{Submitted 2004 September 7.
      Received ;
      in original form}




\begin{abstract}
It has recently been demonstrated that the Shakura-Sunyaev prescription
for the kinematic viscosity in an advection-dominated accretion disk
yields physically reasonable solutions for the structure of the inflow
close to the event horizon. In particular, no violations of relativistic
causality occur at the horizon. This is somewhat surprising considering
the diffusive nature of the angular momentum transport in the
Shakura-Sunyaev scenario, and it is therefore natural to ask whether one
can also obtain acceptable solutions for the disk structure based on the
various alternative models for the viscosity that have been proposed,
including the ``deterministic'' forms. In this paper we perform a
rigorous asymptotic analysis of the structure of an advection-dominated
accretion disk close to the event horizon of a nonrotating black hole
based on three of the alternative prescriptions for the viscosity that
have been suggested in the literature.

We constrain the physical disk model by stipulating that the stress must
vanish at the horizon, which is the fundamental inner boundary condition
imposed by general relativity. Surprisingly, we find that none of the
three alternative viscosity prescriptions yield physically acceptable
disk structures close to the horizon when the zero-torque condition is
applied, whether the flow is in vertical hydrostatic equilibrium or
free-fall. Hence we conclude that the original Shakura-Sunyaev
prescription is the only one proposed so far that is physically
consistent close to the event horizon. We argue that, somewhat
ironically, it is in fact the diffusive nature of the Shakura-Sunyaev
form that is the reason for its success in this application. Our focus
here is on advection-dominated accretion disks, but we expect that our
results will also apply to generalized disks provided that losses of
matter and energy become negligible as the gas approaches the event
horizon.
\end{abstract}


\section{INTRODUCTION}

The advection-dominated accretion flow (ADAF) model remains a popular
scenario for the physical structure of the accretion disks in X-ray
underluminous, radio-loud active galactic nuclei (AGNs), as first
proposed by Narayan \& Yi (1994, 1995). The sub-Eddington accretion
rates in these systems cause the plasma to be rather tenuous, which
greatly reduces the efficiency of two-body radiative processes such as
thermal bremsstrahlung. Consequently, the ratio of the X-ray luminosity
divided by the accretion rate is much lower than that associated with
luminous AGNs, which presumably have near-Eddington accretion rates. In
the standard ADAF scenario, the ions absorb most of the energy
dissipated via viscosity and achieve a nearly virial temperature ($T_i
\sim 10^{12}\,$K), far in excess of the electron temperature ($T_e \sim
10^9\,$K). The energy per unit mass in the ions is therefore comparable
to the gravitational binding energy, and consequently most of the energy
deposited in the disk by viscous dissipation is advected across the
event horizon into the black hole, unless outflows of relativistic
particles are able to significantly cool the disk (Becker, Subramanian,
\& Kazanas 2001; Blandford \& Begelman 1999).

\subsection{Causality and Stress in the Inner Region}

One of the most intriguing unresolved questions related to the structure
of ADAF disks, and indeed, to accretion disks in general, concerns the
behavior of the torque in the inner region, where the material begins to
plunge into the black hole. This issue is of central importance in the
development of computational models because the boundary conditions
applied in the inner region can influence the structure of the entire
disk. A variety of approaches have been applied towards the modeling of
the variation of the torque in the inner regions of accretion disks. For
example, in the context of standard thin-disk accretion, it is usually
supposed that the stress vanishes (and the disk truncates) at the
marginally stable orbit (e.g., Frank, King, \& Raine 1985), although
this suggestion has been contradicted by Watarai \& Mineshige (2003)
based on the results of hydrodynamical simulations. Furthermore, several
authors have argued that the stress must actually vanish at the sonic
point (e.g., Kato 1994; Popham \& Narayan 1992). These models are based
on what one might term the ``hydrodynamic'' causality scenario, in which
the stress is transmitted by subsonic turbulence, and therefore no
torque can exist in the inner, supersonic region of the flow.

The situation in an actual accretion disk is far from clear, and the
angular momentum may actually be transported by some combination of
fluid turbulence, particles, and electromagnetic fields. The velocity of
the viscous transport associated with particles and fields is not
limited to the sound speed, and therefore it is plausible that torques
can be generated even in the supersonic region between the horizon and
the sonic point (e.g., Zimmerman et al. 2004; Reynolds \& Armitage 2001;
Hawley \& Krolik 2001; Agol \& Krolik 2000; Gammie 1999). In these
scenarios, the ``hydrodynamic'' causality restriction must be replaced
with the more fundamental ``relativistic'' causality constraint, which
states that {\it no signal of any kind can propagate faster than the
speed of light}. We shall refer to flows satisfying this requirement as
``deterministic.'' The associated relativistic restriction on the torque
is that {\it it must vanish at the event horizon}, since the horizon
itself cannot support a shear stress (Weinberg 1972). Taken together,
these two related constraints comprise the most conservative and
model-independent statements one can make about the causal and viscous
structure of any accretion disk, and we shall therefore adopt them as
the basis for the analysis presented here.

\subsection{Angular Momentum Transport}

The specific results obtained for the structure of an accretion disk
depend on the detailed prescription employed for the kinematic viscosity,
$\nu$, which establishes the basic connection between the angular velocity
$\Omega$, the torque ${\cal G}$, and the stress (force per unit area)
$\Sigma$ through the expression
\begin{equation}
\Sigma = - \rho \, \nu \, r \, {d\Omega \over dr}
= {{\cal G} \over 4 \pi \, r^2 \, H} \ ,
\label{eq1}
\end{equation}
where $\rho$ is the mass density, $H$ is the half-thickness of the disk,
and the disk is rotating differentially so that $d\Omega/dr < 0$.
Various prescriptions for the physical form of $\nu$ have been proposed
over the years, starting with the ``diffusive'' approach employed by
Shakura \& Sunyaev (1973). In this scenario, the angular momentum
transport is governed by a one-dimensional diffusion equation, which
technically leads to the propagation of an infinitesimal amount of
signal to infinite distances in zero time (Pringle 1981). The apparent
violation of relativistic causality associated with the Shakura-Sunyaev
formulation has stimulated other workers to consider a variety of
alternative, ``deterministic'' forms for the viscosity (e.g., Narayan,
Kato, \& Honma 1997; Yuan et al. 2000). These alternative forms are
important in the context of accretion onto compact objects with solid
surfaces, such as neutron stars and white dwarfs, because in these
environments the diffusive Shakura-Sunyaev approach can result in a
non-causal disk structure.

However, the results are quite different when one considers accretion
onto a black hole. In this case, Becker \& Le (2003) demonstrated
conclusively that despite the diffusive nature of the Shakura-Sunyaev
scenario, there are in fact no relativistic causality violations in the
region close to the event horizon. This is because the propagation of
signals near the horizon is dominated by {\it inward advection} at the
speed of light, which overwhelms the outward, diffusive propagation.
Furthermore, Becker \& Le used the standard conservation equations to
confirm that the Shakura-Sunyaev prescription yields physically
reasonable global solutions for the structure of ADAF disks, whether the
disk is in vertical hydrostatic equilibrium near the horizon or in
free-fall. The fact that the original Shakura-Sunyaev prescription
yields an acceptable disk structure close to the black hole naturally
causes one to ask whether this is also true for the alternative
viscosity formulations suggested by subsequent authors. Our goal in this
paper is to apply the same asymptotic analysis technique employed by
Becker \& Le (2003) in the vicinity of the event horizon to answer the
``existence question'' for the two specific deterministic viscosity
prescriptions considered by Narayan, Kato, \& Honma (1997) and Yuan et
al. (2000). We shall also include the model proposed by Richard \& Zahn
(1999) which, although not deterministic, is nonetheless interesting to
examine from a conceptual viewpoint.

We shall focus mainly on ADAF disks in the present paper because they
appear to be correlated with the presence of outflows in systems
possessing high radio luminosities and low X-ray luminosities. The low
density plasmas in the hot ADAF disks seem to provide ideal
environments for the shock acceleration of relativistic particles that
can escape to power the outflows (Le \& Becker 2004). The apparent
association of ADAF disks with outflows and jets suggests that a
complete understanding of the disk structure is essential in order to
make further progress in the development of global models that
simultaneously account for the accretion of the gas as well as the
production of the outflows in a self-consistent manner. ADAF disks are
also particularly amenable to mathematical modeling because the energy
transport is simplified by the fact that radiative losses are negligible
throughout the flow. Furthermore, it is expected that the escape of
matter and energy due to the acceleration of relativistic particles in
the disk occurs outside the radius of marginal stability, and therefore
the variation of the internal energy density is essentially adiabatic in
the vicinity of the horizon, where the viscous dissipation becomes
negligible (Le \& Becker 2004; Becker et al. 2001). Although our discussion
here centers on ADAF disks around nonrotating black holes, we will argue
that the constraints obtained on the form for the viscosity in the vicinity
of the event horizon also apply to more general flows, provided they do not
radiate strongly near the horizon.

In the standard Shakura-Sunyaev formulation of the viscosity variation
in the disk, the transport of angular momentum is a diffusive process
and the time evolution of the angular velocity $\Omega$ is therefore
governed by a second-order differential equation, requiring two boundary
conditions. One of these conditions is provided by imposing that the
viscous stress must vanish at the event horizon of the black hole, which
is a mandatory requirement of general relativity (Weinberg 1972).
However, two of the alternative viscosity prescriptions examined here
are ``deterministic'' in the sense that they result in first-order
differential equations for the time evolution of $\Omega$. In these
cases, obviously one less boundary condition is required in order to
specify the global flow solution. We argue that regardless of which
formulation is adopted for the viscosity, the fundamental zero-stress
boundary condition at the event horizon must be retained for consistency
with general relativity. Further discussion of this point is provided in
\S~5.1.

The remainder of the paper is organized as follows. In \S~2 we
briefly review the fundamental equations governing the structure of
one-dimensional ADAF disks. In \S~3 the causal structure of the
viscous transport in the disk is discussed in the context of the four
viscosity prescriptions of interest here. The associated existence
conditions are derived in \S~4 under the assumption of either
vertical hydrostatic equilibrium or free-fall in the inner region of the
disk. The implications of our results for the physical variation of the
viscosity in advection-dominated black-hole accretion disks are
discussed in \S~5.

\section{FUNDAMENTAL EQUATIONS}

The approach to the modeling of the disk structure is simplified
considerably if the effects of general relativity are incorporated in an
approximate manner by expressing the gravitational potential per unit
mass using the pseudo-Newtonian form (Paczy\'nski \& Wiita 1980)
\begeq
\Phi(r) \equiv {- G \, M \over r - \rs} \ ,
\label{eq2}
\fineq
where $\rs = 2GM/c^2$ is the Schwarzschild radius for a black hole of
mass $M$. Becker \& Le (2003) confirmed that this potential reproduces
perfectly the motions of particles falling freely in the Schwarzschild
metric close to the event horizon. It also correctly predicts the
location of the event horizon, the radius of marginal stability, and the
radius of the marginally bound orbit (for a complete discussion, see
Paczy\'nski \& Wiita 1980). This potential, while providing a good
approximation of the effects of general relativity, is also rather
convenient mathematically because it facilitates a semi-classical
approach to the problem that simplifies the analysis considerably. Due
to these advantages, the pseudo-Newtonian potential has been utilized by
many authors in their studies of the accretion of gas onto Schwarzschild
black holes (e.g., Matsumoto et al. 1984; Abramowicz et al. 1988; Chen
et al. 1997; Narayan, Kato, \& Honma 1997; Hawley \& Krolik 2001, 2002;
Yuan 1999; Yuan et al. 2000; Reynolds \& Armitage 2001).

\subsection{Energy and Momentum Conservation}

Since ADAF disks are radiatively inefficient, the vertical height can
become comparable to the radius and therefore it is appropriate to
utilize the vertically-averaged ``slim disk'' equations first discussed
by Abramowicz et al. (1988). These equations were adopted by Narayan \&
Yi (1994) in their study of self-similar ADAF solutions, and also by
Narayan, Kato, \& Honma (1997) in their simulations of transonic flows.
In the slim-disk approximation, the inertial ($v\,dv/dr$) and pressure
gradient ($\rho^{-1} \, dP/dr$) terms are retained in the radial
momentum equation, so that in a steady state the radial acceleration
rate in the frame of the accreting gas is given by
\begeq
{Dv \over Dt} \equiv
- \vr \, {d\vr \over dr} = {1\over \rho} {dP \over dr}
+ {d\Phi \over dr} - r \, \Omega^2 \ ,
\label{eq2.1.1}
\fineq
where
the radial velocity $v$ is defined to be
positive for inflow. The escape of energy from the disk is assumed to be
negligible in the ADAF approximation, and therefore the rate of change
of the internal energy density $U$ in the frame of the gas can be
written as
\begeq
{DU \over Dt} \equiv
- \vr \, {dU \over dr} = - \gamma \, {U \vr \over \rho} \, {d\rho \over
dr} + \dot U_{\rm viscous} \ ,
\label{eq2.1.2}
\fineq
where
\begeq
\dot U_{\rm viscous} = - \, {{\cal G} \over 4 \pi r H} {d\Omega \over dr}
= - \, r \, \Sigma \, {d\Omega \over dr}
\label{eq2.1.3}
\fineq
is the viscous energy dissipation rate per unit volume and $\gamma$
denotes the specific heats ratio. In a steady-state situation,
the radial variation of the angular velocity $\Omega$ is determined by
(see eq.~[\ref{eq1}])
\begeq
{d\Omega \over dr} = - {{\cal G} \over 4 \pi \, r^3 \, H \, \rho \, \nu} \ .
\label{eq2.1.4}
\fineq

\subsection{Transport Rates}

In one-dimensional, stationary, advection-dominated disks, three
integrals of the flow can be identified. The first is the accretion
rate,
\begin{equation}
\dot M = 4 \pi \, r \, H \, \rho \, v = \rm constant \ ,
\label{eq3}
\end{equation}
and the second is the angular momentum transport rate,
\begin{equation}
\dot J = \dot M \, r^2 \, \Omega - {\cal G} = \rm constant \ .
\label{eq4}
\end{equation}
The third conserved quantity is the energy transport rate,
\begin{equation}
\dot E = - {\cal G} \, \Omega + \dot M \left({1 \over 2} \, v^2
+ {1\over 2} \, w^2 + {P+U \over \rho} + \Phi \right)
= \rm constant
\ ,
\label{eq5}
\end{equation}
where $U = P/(\gamma-1)$ is the internal energy density and
$w=r\,\Omega$ is the azimuthal velocity. The constancy of $\dot E$ can
be demonstrated explicitly by combining equations~(\ref{eq2.1.1}),
(\ref{eq2.1.2}), (\ref{eq2.1.3}), (\ref{eq3}), and (\ref{eq4}). Despite
the classical appearance of the conservation equations for $\dot M$,
$\dot J$, and $\dot E$, it is important to bear in mind that close to
the horizon, $v$ and $w$ are actually more correctly interpreted as the
radially and azimuthal components of the four-velocity, respectively
(see Becker \& Le 2003).

The vanishing of the stress and the torque at the horizon, combined
with equation~(\ref{eq4}), together imply that
\begin{equation}
\lim_{r \to \rs} \Omega(r) = {\dot J \over \dot M \, \rs^2}
\equiv \Omega_0 \ ,
\label{eq6}
\end{equation}
and therefore $\Omega$ achieves a finite value at the horizon. By
eliminating the torque ${\cal G}$ between equations~(\ref{eq4}) and
(\ref{eq5}), we can reexpress the energy transport rate as
\begin{equation}
\dot E = \dot J \, \Omega + \dot M \left({1 \over 2} \, v^2
- {1\over 2} \, r^2 \, \Omega^2 + {a^2 \over \gamma-1}
+ \Phi \right)
\ ,
\label{eq7}
\end{equation}
where 
\begin{equation}
a \equiv \left(\gamma P \over \rho \right)^{1/2}
\label{eq8}
\end{equation}
denotes the adiabatic sound speed.

Since the angular velocity $\Omega$ approaches a finite value at the
horizon and the flow must be supersonic there, it follows from
equation~(\ref{eq7}) that
\begin{equation}
v(r) \to v_{\rm ff}(r) \equiv \left(GM \over r-\rs \right)^{1/2}
\ , \ \ \ \ \ \ \ r \to \rs \ ,
\label{eq9}
\end{equation}
where $v_{\rm ff}$ denotes the free-fall velocity in the
pseudo-Newtonian potential. Note that the divergence of $v_{\rm ff}$
implies that it is not a Newtonian velocity, but rather a four-velocity,
as established by Becker \& Le (2003). Equation~(\ref{eq9}) clearly indicates
that the inflow approaches free-fall close to the event horizon, and
consequently the radial velocity $v^{\hat r}$ approaches the speed
of light (Shapiro \& Teukolsky 1983).

\subsection{Specific Angular Momentum}

Next we shall focus on the asymptotic variation of the angular velocity
$\Omega$ and the specific angular momentum $\ell \equiv r^2 \, \Omega$
in the limit $r \to \rs$. Equation (\ref{eq4}) can be rewritten in terms
of $\ell$ as
\begin{equation}
\ell - \ell_0 = {{\cal G} \over \dot M}
\ ,
\label{eq10}
\end{equation}
where $\ell_0 \equiv \dot J/\dot M = \rs^2 \,\Omega_0$ is the accreted
specific angular momentum. Since the torque ${\cal G}$ is positive for
$r > \rs$ and it vanishes at the horizon ($r = \rs$) , it follows from
equation~(\ref{eq10}) that
\begin{equation}
\lim_{r \to \rs} {d \ell \over dr} \ge 0 \ .
\label{eq11}
\end{equation}
Based on the relationship between $\Omega$ and $\ell$, we find that
\begin{equation}
{d \ell \over dr} = 2 \, r \, \Omega + r^2 \, {d \Omega \over dr}
\ .
\label{eq12}
\end{equation}
We can express the variation of $\Omega(r)$ in the neighborhood of the
event horizon using the general form
\begin{equation}
\Omega(r) \doteq \Omega_0 - A \, (r-\rs)^q \ ,
\label{eq13}
\end{equation}
where $A$ and $q$ are positive constants and the symbol ``$\doteq$''
will be used to denote asymptotic equality at the horizon. This is the
simplest form satisfying the conditions $\Omega(r) \to \Omega_0$ and
$d\Omega/dr \le 0$ as $ r \to \rs$. From a slightly more technical point
of view, the right-hand side of equation~(\ref{eq13}) represents the
first two terms in the Frobenius expansion of $\Omega(r)$ around $r =
\rs$, in which case $q$ is the exponent of the solution (Boyce \&
DiPrima 1977). The values of the constants $A$ and $q$ are determined
through analysis of the conservation equations.

Combining equations~(\ref{eq12}) and (\ref{eq13}) yields for the
asymptotic behavior of the specific angular momentum
\begin{equation}
{d \ell \over dr} \doteq 2 \, r \, \Omega_0 - 2 A \, r \,
(r-\rs)^q - q A \, r^2 \, (r-\rs)^{q-1} \ .
\label{eq14}
\end{equation}
This result implies that we must have $q \ge 1$ in order to avoid
divergence of $d\ell/dr$ to negative infinity at the horizon, which is
unacceptable according to equation~(\ref{eq11}). The restriction on $q$
in turn implies that $d\ell/dr$ {\it must be equal to zero or a finite
positive value at the horizon}. Following Becker \& Le (2003), we shall
express the variation of the specific angular momentum in the vicinity
of the event horizon using the general form
\begeq
\ell(r) \asym \ell_0 + B \, (r-\rs)^\beta \ , \ \ \ \ \ \beta \ge 1 \ ,
\label{eq15}
\fineq
where $B$ is a positive constant. This equation represents the first two
terms of the Frobenius expansion for $\ell(r)$ about $r = \rs$, with
$\beta$ denoting the exponent of the solution (cf. equation~\ref{eq13}).
The restriction $\beta \ge 1$ is required in order to ensure that
$d\ell/dr$ is equal to zero or a finite positive quantity at the event
horizon, as established above. Different values of $\beta$ will be
obtained depending on the viscosity model employed, as discussed in
\S~4. It should be emphasized that any viscosity model that yields
$\beta < 1$ fails to satisfy the basic existence condition, and must be
rejected. While Paczy\'nski \& Wiita (1980) imposed
equation~(\ref{eq15}) as an ad hoc expression for the global variation
of the specific angular momentum $\ell(r)$, we shall utilize it only in
the asymptotic limit, where it is fully consistent with the conservation
equations.

Equations~(\ref{eq12}) and (\ref{eq15}) together imply that the radial
derivative of $\Omega$ at the horizon is given by
\begeq
{d \Omega \over dr} \bigg|_{r = \rs}
= \cases{
- 2 \, \ell_0 / \rs^3 \ , & $\beta > 1$ \ , \cr
\phantom{space} \cr
(B \, \rs - 2 \, \ell_0) / \rs^3 \ , & $\beta = 1$ \ . \cr
}
\label{eq16}
\fineq
Hence $d\Omega/dr$ vanishes at the horizon only in the special case
$\beta=1$ and $B=2\,\ell_0/\rs$. These conditions are not satisfied in
the situations of interest here, and therefore $d\Omega/dr$ has a finite
(negative) value at the horizon. By combining equations~(\ref{eq1}),
(\ref{eq10}), and (\ref{eq15}), we obtain the fundamental asymptotic
relation
\begeq
B \, (r-\rs)^\beta \asym - \, {r^2 \, \nu \over \vr} \, {d \Omega \over dr}
\ .
\label{eq17}
\fineq
This expression will prove useful in establishing the existence
conditions for inflows subject to a variety of prescriptions for the
kinematic viscosity $\nu$. The procedure will be to compute the value of
$\beta$ for each viscosity model by balancing the powers of $(r - \rs)$
on the two sides of equation~(\ref{eq17}), and then to compare the result
with the existence condition $\beta \ge 1$. Satisfaction of this condition
ensures that the variation of the specific angular momentum and its derivative
are physically acceptable close to the event horizon, and also that the
torque vanishes there as required. 

\subsection{Entropy Function}

Following Becker \& Le (2003), we will adopt the ``perfect ADAF''
approximation, and consequently the escape of energy from the disk will
be ignored for the moment. Since the stress $\Sigma$ vanishes as $r \to
\rs$, it follows that the flow approaches a purely adiabatic behavior
($U \propto \rho^\gamma$) in the vicinity of the event horizon (see
eq.~[\ref{eq2.1.3}]). Furthermore, if the gas is in local thermodynamic
equilibrium, then the viscous heating is a quasi-static process, and in
this case the flow is isentropic wherever the dissipation vanishes.

In our analysis of the flow structure close to the horizon, we shall find
it convenient to introduce the ``entropy function,''
\begeq
K(r) \equiv r \, H \, \vr \, a^{2/(\gamma-1)}
\ .
\label{eq20}
\fineq
To understand the physical significance of $K$, we can combine
equations~(\ref{eq3}), (\ref{eq8}), and (\ref{eq20}) to show that
\begeq
K^{\gamma-1} \propto {U \over \rho^\gamma} \ .
\label{eq21}
\fineq
This result establishes that $K$ has a constant value near the event
horizon because the viscous dissipation rate vanishes there and the flow
becomes adiabatic. It is interesting to note that if the gas is in local
thermodynamic equilibrium, then we can use equation~(\ref{eq21}) to show
that the value of $K$ is connected to the entropy per particle $S$ by
(Reif 1965)
\begeq
S = k \, \ln K + c_0 \ ,
\label{eq22}
\fineq
where $c_0$ is a constant that is independent of the state of the gas,
but may depend upon its composition. The constancy of $K$ near the
horizon will allow us to obtain a convenient expression for the
asymptotic variation of the adiabatic sound speed $a$ in terms of
$r$ and $\vr$ for disks in hydrostatic equilibrium or free-fall,
as discussed in \S~4.

\section{VISCOUS TRANSPORT AND CAUSALITY}

The various prescriptions for the kinematic viscosity $\nu$ considered by
Shakura \& Sunyaev (1973), Narayan, Kato, \& Honma (1997), Yuan et al.
(2000), and Richard \& Zahn (1999) have very different implications for
the (relativistic) causal structure of the angular momentum transport
occurring in an accretion disk. The distinctions between the different
approaches can be understood clearly by analyzing the equation
describing the time-dependent transport of angular momentum in the disk.
Following Becker \& Le (2003) and Blandford \& Begelman (1999), we write
\begeq
{\partial \over \partial t} \left(\mu r^2 \Omega \right)
= {\partial \over \partial r} \left(\mu r^2 \Omega v - {\cal G}
\right) \ ,
\label{eq23}
\fineq
where
\begeq
\mu \equiv 4 \pi r H \rho = {\dot M \over v}
\label{eq24}
\fineq
represents the mass per unit radius in the disk. The corresponding conservation
equation for $\mu$ is given by
\begeq
{\partial \mu \over \partial t}
= {\partial \over \partial r} \left(\mu v \right) \ .
\label{eq25}
\fineq
By combining equations~(\ref{eq1}) and (\ref{eq24}), we find that the
torque ${\cal G}$ can be expressed in terms of $\mu$, $\nu$, and
$\Omega$ as
\begeq
{\cal G} = - \, \mu r^2 \nu \, {\partial \Omega \over \partial r} \ .
\label{eq26}
\fineq
Utilizing equations~(\ref{eq23}), (\ref{eq25}), and (\ref{eq26}), we can
show that the time derivative of $\Omega$ is given by
\begeq
{\partial \Omega \over \partial t} = {1 \over \mu r^2} \, {\partial \over \partial r}
\left(\mu r^2 \Omega v + \mu r^2 \nu {\partial \Omega \over \partial r}\right)
- {\Omega \over \mu} \, {\partial \over \partial r} \left(\mu v\right) \ .
\label{eq27}
\fineq
The nature of this equation depends on the functional form assumed for the
variation of $\nu$, which determines whether equation~(\ref{eq27}) is first-
or second-order in $\Omega$, and whether it is linear or nonlinear. We shall
discuss below the implications of each of the four viscosity prescriptions
of interest here.

\subsection{Shakura \& Sunyaev}

Shakura \& Sunyaev (1973) were the first to suggest that the variation
of the kinematic viscosity in an accretion disk can be approximated using
the form
\begeq
\nu =\alpha \, a \, H \ ,
\label{eq28}
\fineq
with the corresponding shear stress (see eq.~[\ref{eq1}])
\begeq
\Sigma = - \alpha \, a \, H \, \rho \, r \,
{d\Omega \over d r} \ .
\label{eq29}
\fineq
where $\alpha$ is a positive constant of order unity. The formulation
is based on the idea that the viscosity may be expected to be roughly
proportional to the product of the turbulent velocity (usually some
fraction of the sound speed) and the largest scale of the turbulent
eddies (the disk height $H$). This fundamental prescription should hold
whether the disk is hydrostatic or in free-fall. Combining
equations~(\ref{eq27}) and (\ref{eq28}) yields in the case of the
Shakura-Sunyaev viscosity
\begeq
{\partial \Omega \over \partial t} = {1 \over \mu r^2} \, {\partial
\over \partial r} \left(\mu r^2 \Omega v + \alpha a H \mu r^2
{\partial \Omega \over \partial r}\right) - {\Omega \over \mu} \,
{\partial \over \partial r} \left(\mu v\right) \ .
\label{eq31}
\fineq
This second-order equation in $\Omega$ has a diffusive character,
because the angular momentum is always transported in the direction
opposed to the radial gradient of $\Omega$ (Pringle 1981). We can obtain
additional insight on this point by focusing on the time evolution of an
initially localized component of the angular momentum distribution,
represented by a $\delta$-function at some arbitrary radius $r=r_0$ at
an arbitrary time $t=t_0$. As time advances, the distribution will
initially spread in radius in an approximately Gaussian manner, implying
the propagation of an infinitesimal portion of the angular momentum to
infinite distance in a finite time.

The diffusive nature of the transport in this case has led a number of
authors to point out the potential for violations of relativistic
causality in accretion disks (e.g., Kato 1994; Narayan 1992). However,
as discussed by Becker \& Le (2003), this phenomenon has a negligible
effect on the structure of an accretion disk in the outer, subsonic
region, because the {\it mean} transport velocity for the angular
momentum remains small despite the fact that an infinitesimal amount of
angular momentum is transported to an infinite distance in zero time.
Moreover, in the context of accretion onto a nonrotating black hole,
they demonstrated though an explicit derivation of the relevant
Fokker-Planck coefficients that there are no relativistic causality
violations near the event horizon associated with the Shakura-Sunyaev
viscosity prescription. This is due to the fact that signals originating
near the horizon are simply advected into the black hole at the speed of
light, in agreement with general relativity. In a broader context,
however, it is important to note that the causality issue remains a
central consideration in situations involving the accretion of matter
onto the solid surfaces of neutron stars and white dwarfs.

\subsection{Narayan, Kato, \& Honma}

The relativistic causality problem associated with the original
Shakura-Sunyaev prescription for the variation of $\nu$ has motivated
several authors to consider a variety of ``deterministic'' alternatives
to equation~(\ref{eq28}). These all involve the replacement of
equation~(\ref{eq29}) for the stress with a new expression that does not
contain $d\Omega/dr$, thereby rendering equation~(\ref{eq27})
first-order in $\Omega$. Since the transport of angular momentum is no
longer a diffusive process, these alternatives satisfy the relativistic
causality constraint. We note, however, that they may still fail the
``hydrodynamic'' causality test discussed in \S~1.1, although this is
not necessarily a problem if the stress is transmitted by particles
and/or fields rather than by fluid turbulence. For example, Narayan,
Kato, \& Honma (1997) considered the alternative form for the viscous
stress
\begin{equation}
\Sigma = - \, \alpha \, P \, {d\ln\OmegaK \over d\ln r} \ ,
\label{eq33}
\end{equation}
with the associated kinematic viscosity given by
\begin{equation}
\nu = {\alpha \, a^2 \over \gamma \, \OmegaK} \,
\left(d\OmegaK \over dr \right)
\left(d\Omega \over dr \right)^{-1}
\ ,
\label{eq32}
\end{equation}
where $\OmegaK(r)$ represents the Keplerian angular velocity of matter
in a circular orbit at radius $r$. In the case of the pseudo-Newtonian
potential given by equation~(\ref{eq2}), we obtain
\begeq
\OmegaK^2(r) \equiv {GM \over r \, (r-\rs)^2}
= {1 \over r} \, {d\Phi \over dr} \ .
\label{eq30}
\fineq
Substituting equation~(\ref{eq32}) into equation~(\ref{eq27}) yields
\begeq
{\partial \Omega \over \partial t} = {1 \over \mu r^2} \, {\partial
\over \partial r} \left(\mu r^2 \Omega v + {\alpha a^2 \mu r^2 \over
\gamma \, \OmegaK} {d \OmegaK \over d r}\right) - {\Omega \over \mu} \,
{\partial \over \partial r} \left(\mu v\right) \ .
\label{eq34}
\fineq
This first-order equation for $\Omega$ has no diffusive character, and
therefore the propagation of perturbations in the angular momentum
distribution will occur in a deterministic manner. However, despite this
apparent advantage, we shall see in \S~4 that the prescription for
$\nu$ given by equation~(\ref{eq32}) implies an unphysical structure in
the inner region of an accretion flow around a nonrotating black hole.

\subsection{Yuan et al.}

Another ``deterministic'' form for the viscosity was suggested by Yuan
et al. (2000; see also Matsumoto et al. 1984 and Shakura \& Sunyaev 1973),
who proposed that
\begin{equation}
\Sigma = \alpha \, P \ ,
\label{eq36}
\end{equation}
which yields for the viscosity
\begin{equation}
\nu = - \, {\alpha \, a^2 \over \gamma \, r} \, \left(d\Omega \over dr
\right)^{-1} \ .
\label{eq35}
\end{equation}
Substitution into equation~(\ref{eq27}) now yields
\begeq
{\partial \Omega \over \partial t} = {1 \over \mu r^2} \, {\partial
\over \partial r} \left(\mu r^2 \Omega v - \alpha \gamma^{-1} a^2 \mu r\right)
- {\Omega \over \mu} \, {\partial \over \partial r} \left(\mu v\right) \ .
\label{eq37}
\fineq
As in the previous example, we again obtain a first-order equation in
$\Omega$ that does not have any diffusive character. While this form for
$\nu$ does result in angular momentum transport that satisfies the
relativistic causality constraint, we will nonetheless find that
equation~(\ref{eq35}) produces an unphysical structure when applied in
the inner region of an accretion flow.

\subsection{Richard \& Zahn}

Based on laboratory studies of differentially rotating flows in the
Couette-Taylor experiment, Richard \& Zahn (1999) proposed that the
kinematic viscosity in accretion disks scales as
\begin{equation}
\nu = - \alpha \, r^3 {d \Omega \over dr} \ ,
\label{eq38}
\end{equation}
with the associated stress
\begin{equation}
\Sigma = \alpha \, r^4 \rho \left(d \Omega \over dr\right)^2 \ .
\label{eq39}
\end{equation}
Combining equations~(\ref{eq27}) and (\ref{eq38}) yields for the
Richard \& Zahn scenario
\begeq
{\partial \Omega \over \partial t} = {1 \over \mu r^2} \,
{\partial \over \partial r}
\left[\mu r^2 \Omega v - \alpha \mu r^5 \left(\partial \Omega \over
\partial r\right)^2\right] - {\Omega \over \mu} \,
{\partial \over \partial r}
\left(\mu v\right) \ .
\label{eq40}
\fineq
This expression for $\partial \Omega/\partial t$ is quite different from
those obtained for the other three viscosity prescriptions, because it
is a nonlinear diffusion equation. From a physical point of view, the
diffusive character of equation~(\ref{eq40}) implies a violation of the
relativistic causality requirement. Furthermore, the nonlinear nature of
equation~(\ref{eq40}) results in the propagation of angular momentum in
the {\it outward} direction {\it regardless} of the sign of the gradient
$d\Omega/dr$. Despite this unusual property, it is possible to model
accretion disks using equation~(\ref{eq38}) for $\nu$ because
$d\Omega/dr$ is generally negative in disks, which is consistent with
angular momentum transport in the outward direction. However, we shall
see that the structure of the innermost region of an accretion flow
computed using equation~(\ref{eq38}) is unphysical.

\section{ASYMPTOTIC BEHAVIOR IN ADVECTION-DOMINATED DISKS}

Many of the theoretical models for advection-dominated accretion disks
appearing in the literature have been based on the assumption of
vertical hydrostatic equilibrium. While this assumption is valid in the
outer, subsonic region, it is clear that close to the horizon, where the
inflow becomes supersonic, sound waves do not have enough time to
maintain hydrostatic equilibrium before the gas enters the black hole.
One therefore expects that in the inner region, the disk will have a
free-fall structure, with the disk half-thickness $H$ proportional to
the radius $r$. Nonetheless, the hydrostatic assumption has been used to
model the structure of accretion flows all the way in to the event
horizon. The popularity of the hydrostatic model, combined with the
physical plausibility of the free-fall model, motivates us to consider
both possibilities here. Consequently, our goal in this section is to
determine whether or not the ``existence condition'' $\beta \ge 1$ (see
equation~\ref{eq15}) is satisfied for each of the four viscosity
prescriptions of interest, in both hydrostatic and free-fall disks. In
each of our calculations we shall make use of the asymptotic relations
(see equations~\ref{eq9} and \ref{eq30})
\begeq
v \propto (r-\rs)^{-1/2} \ , \ \ \ \ 
\OmegaK \propto (r-\rs)^{-1} \ , \ \ \ \ \ 
r \to \rs \ ,
\label{eq41}
\fineq
which are valid in general.

\subsection{Asymptotic Behavior in Hydrostatic Disks}

The results discussed in \S\S~2 and 3 apply to all
advection-dominated disks in the pseudo-Newtonian potential. In this
section, we shall specialize to the case of ADAF disks that maintain
vertical hydrostatic equilibrium all the way in to the event horizon.
While this is not necessarily justified on physical grounds, it
nonetheless provides a useful basis for comparison with several
published models based on this assumption.

For accretion disks in vertical hydrostatic equilibrium, the vertical
half-thickness of the disk $H$ is given by the standard relation
(Abramowicz et al. 1988)
\begeq
H(r) = {b_0 \, a \over \OmegaK} \ ,
\label{eq42}
\fineq
where $b_0$ is a dimensionless constant of order unity that depends
on the details of the vertical averaging, and $\OmegaK(r)$ is given
by equation~(\ref{eq30}). By combining equations~(\ref{eq20}) and
(\ref{eq42}), we find that the entropy function in a hydrostatic disk
is given by
\begeq
K = K_{\rm eq} \equiv {b_0 \, r \, \vr \over \OmegaK} \,
a^{\gamma+1 \over \gamma-1} \ .
\label{eq43}
\fineq
Close to the event horizon, where dissipation becomes unimportant and
$K$ is essentially constant, the variation of the adiabatic sound speed
$a$ can therefore be expressed as
\begeq
a \propto \left(\OmegaK \over \vr\right)^{\gamma-1 \over \gamma+1}
\ , \ \ \ \ r \to \rs \ .
\label{eq44}
\fineq
By combining this result with equations~(\ref{eq41}), we find that the
explicit asymptotic radial dependence of the sound speed in a
hydrostatic disk is given by
\begeq
a^2 \propto (r - \rs)^{1-\gamma \over 1+\gamma}
\ , \ \ \ \ r \to \rs \ .
\label{eq45}
\fineq
Note that $a \to \infty$ as $r \to \rs$ due to the effect of adiabatic
compression as the gas flows towards the horizon. This expression will
prove useful when we analyze the existence conditions for hydrostatic
accretion disks in the vicinity of the event horizon.

\subsection{Asymptotic Behavior in Freely-Falling Disks}

ADAF disks are likely to maintain vertical hydrostatic structure in the
outer, subsonic region. However, close to the horizon, the inflow must
become supersonic and therefore sound waves will not be able to maintain
hydrostatic equilibrium (e.g., Matsumoto et al. 1984). The transition between
the hydrostatic, subsonic outer region and the freely-falling, supersonic
inner region can be modeled using the prescription suggested by Abramowicz,
Lanza, \& Percival (1997). However, in the present paper, we are interested
only in the asymptotic behavior near the event horizon. Hence, based on
equation~(18) from Abramowicz et al. (1997), we can express the variation
of the disk half-thickness $H$ using
\begeq
H(r) = d_0 \, r \ ,
\label{eq46}
\fineq
where $d_0$ is a dimensionless constant of order unity. By utilizing
equations~(\ref{eq20}) and (\ref{eq46}), we conclude that the entropy
function in a freely-falling disk is given by
\begeq
K = K_{\rm ff} \equiv d_0 \, r^2 \, \vr \, a^{2 \over \gamma-1} \ .
\label{eq47}
\fineq
The variation of the sound speed close to the event horizon, where $K$
is essentially constant, can be expressed in this case as
\begeq
a \propto v^{(1-\gamma)/2} \ , \ \ \ \ r \to \rs \ .
\label{eq48}
\fineq
The corresponding asymptotic radial dependence of the sound speed close
to the horizon in a freely-falling disk is obtained by combining
equations~(\ref{eq41}) and (\ref{eq48}), which yields
\begeq
a^2 \propto (r - \rs)^{(\gamma - 1)/2}
\ , \ \ \ \ r \to \rs \ .
\label{eq49}
\fineq
Note that $a \to 0$ as $r \to \rs$, in contrast with the hydrostatic
case, where we found that the sound speed diverges at the horizon (cf.
equation~\ref{eq45}). It is interesting to consider how this alternative
central inflow condition will affect our conclusions regarding the
physical applicability of the various viscosity prescriptions in the
region close to the event horizon. We shall treat each of the viscosity
prescriptions separately below, for both freely-falling and hydrostatic
disks.

\subsection{Shakura \& Sunyaev}

In a Shakura-Sunyaev (1973) disk the viscosity and stress are given by
\begeq
\nu = \alpha \, a \, H \ , \ \ \ \ \ 
\Sigma = - \alpha \, a \, H \, \rho \, r \,
{d\Omega \over d r} \ ,
\label{eq50}
\fineq
which can be combined with equation~(\ref{eq17}) to obtain the
asymptotic relation
\begin{equation}
B \, (r - \rs)^\beta \asym - \, {\alpha \, a \, H \, r^2 \over v}
\, {d \Omega \over dr} \ .
\label{eq50b}
\end{equation}
If the disk is in vertical hydrostatic equilibrium, then we can substitute
for $H$ using equation~(\ref{eq42}), which yields
\begin{equation}
B \, (r - \rs)^\beta \asym - \, {\alpha \, b_0 \, a^2 \, r^2 \over v
\, \OmegaK} \, {d \Omega \over dr} \ .
\label{eq51}
\end{equation}
Utilizing the general asymptotic relations given by equations~(\ref{eq41})
along with equation~(\ref{eq45}), we conclude that in order to balance the
exponents of $(r-\rs)$ on both sides of equation~(\ref{eq51}) in a
{\it hydrostatic} Shakura-Sunyaev disk, we must have
\begin{equation}
\beta = {\gamma + 5 \over 2 \, (\gamma+1)} \ \ \ \ \ \ \ 
{\rm (hydrostatic)} \ ,
\label{eq52}
\end{equation}

The asymptotic expression given by equation~(\ref{eq50b}) is valid in
both hydrostatic and freely-falling disks subject to the Shakura-Sunyaev
viscosity prescription. Hence by combining equations~(\ref{eq41}),
(\ref{eq46}), and ({\ref{eq50b}) and balancing powers of $(r-\rs)$
in the resulting expression, we find that in a {\it freely-falling}
Shakura-Sunyaev disk the value of $\beta$ is given by
\begin{equation}
\beta = {\gamma + 1 \over 4} \ \ \ \ \ \ \ 
\phantom{sss} {\rm (free-fall)} \ .
\label{eq53}
\end{equation}
This result is slightly different from equation~(70) in Becker \& Le
(2003), who found that $\beta = 1 + \gamma/2$. The difference between
the two results arises because we have expressed the viscosity in the
freely-falling disk using $\nu \propto a H$, whereas Becker \& Le (2003)
used $\nu \propto a^2/\OmegaK$. It is not obvious which form is the most
appropriate to use in the supersonic region, because either one can be
viewed as the fundamental definition of the viscosity following the
phenomenological arguments given by Shakura \& Sunyaev (1973). In any
event, the difference between the two forms is negligible for the
present considerations since in either case we find that $\beta > 1$.

We therefore conclude that $\beta > 1$ for all values of $\gamma$ in both
equations~(\ref{eq52}) and (\ref{eq53}), and consequently it follows that
the Shakura-Sunyaev prescription for $\nu$ satisfies the existence
condition $\beta \ge 1$ required by equation~(\ref{eq15}) {\it in both
freely-falling and hydrostatic disks}. Moreover, Becker \& Le have also
established conclusively that the Shakura-Sunyaev formulation for the
viscosity is consistent with general relativistic causality requirements
close to the event horizon.

\subsection{Narayan, Kato, \& Honma}

Next we examine the implications of the ``deterministic'' viscosity
prescription considered by Narayan, Kato, \& Honma (1997), for which we
have
\begin{equation}
\nu = {\alpha \, a^2 \over \gamma \, \OmegaK} \,
\left(d\OmegaK \over dr \right)
\left(d\Omega \over dr \right)^{-1}
\ , \ \ \ \ \ \ \ 
\Sigma = - \, \alpha \, P \, {d\ln\OmegaK \over d\ln r} \ .
\label{eq54}
\end{equation}
The angular momentum transport in this case is non-diffusive, and
therefore the relativistic causality constraint is satisfied. Combining
equations~(\ref{eq17}) and (\ref{eq54}), we conclude that
\begin{equation}
B \, (r - \rs)^\beta \asym - \, {\alpha \, r \, a^2 \over \gamma \, v} \,
{d\ln\OmegaK \over d\ln r} \ .
\label{eq55}
\end{equation}
Incorporating the asymptotic behaviors for $v$, $\OmegaK$, and $a$ close to
the horizon given by equations~(\ref{eq41}) and (\ref{eq45}), we now find
that the in a {\it hydrostatic} disk described by the Narayan et al. viscosity,
the exponents of $(r-\rs)$ on the two sides of equation~(\ref{eq55})
balance if
\begin{equation}
\beta = {1 - 3 \, \gamma \over 2 \, (\gamma+1)} \ \ \ \ \ \ \ 
{\rm (hydrostatic)} \ .
\label{eq56}
\end{equation}
Since this quantity is negative for all physically reasonable values of
$\gamma$, it follows that this prescription fails to satisfy the
existence condition $\beta \ge 1$. Hence we conclude that the
alternative, ``deterministic,'' prescription for the viscosity
considered by Narayan, Kato, \& Honma (1997) is unphysical close to the
event horizon in a hydrostatic disk, in apparent contradiction to the
global solutions for the disk structure presented in their paper. This
probably reflects the fact that Narayan et al. did not use their
equation~(2.19) when treating this prescription for the viscosity
variation, which stipulates that the torque vanishes at the horizon.
However, we argue that in fact one does not have the freedom to ignore
this essential physical boundary condition, as discussed in \S~1.1 and
\S~1.2.

Equation~(\ref{eq55}) also applies in the free-fall case, and we can
therefore combine it with equations~(\ref{eq41}) and (\ref{eq49}) to
conclude that in a {\it freely-falling} Narayan et al. disk, the
counterpart of equation~(\ref{eq56}) is given by
\begin{equation}
\beta = {\gamma \over 2} - 1 \ \ \ \ \ \ \ 
\phantom{sss} {\rm (free-fall)} \ .
\label{eq57}
\end{equation}
Hence we once again find that $\beta < 0$ for $4/3 < \gamma < 5/3$, and
therefore we conclude that the deterministic prescription for the
viscosity examined by Narayan, Kato \& Honma (1997) is unphysical close
to the event horizon for both freely-falling and hydrostatic disks.

\subsection{Yuan et al.}

The deterministic prescription suggested by Yuan et al. (2000) gives
the results
\begin{equation}
\nu = - \, {\alpha \, a^2 \over \gamma \, r} \, \left(d\Omega \over dr
\right)^{-1} \ , \ \ \ \ \ \ \ 
\Sigma = \alpha \, P \ .
\label{eq58}
\end{equation}
In this case, the angular momentum transport once again satisfies the
relativistic causality condition since it is non-diffusive in nature.
Combining equations~(\ref{eq17}) and (\ref{eq58}), we now obtain the
asymptotic relation
\begin{equation}
B \, (r - \rs)^\beta \asym {\alpha \, r \, a^2 \over \gamma \, v} \ .
\label{eq59}
\end{equation}
Utilizing equations~(\ref{eq41}) and (\ref{eq45}) to describe the
asymptotic behaviors of $v$ and $a$ close to the horizon, we find that
in a {\it hydrostatic} disk described by the Yuan et al. viscosity, the
exponents of $(r-\rs)$ on the two sides of equation~(\ref{eq59}) balance
if
\begin{equation}
\beta = {3 - \gamma \over 2 \, (\gamma + 1)} \ \ \ \ \ \ \ 
{\rm (hydrostatic)} \ .
\label{eq60}
\end{equation}
Note that $\beta < 1$ for $4/3 < \gamma < 5/3$, which fails to satisfy
the existence condition $\beta \ge 1$. Therefore this viscosity prescription
does not yield a physically acceptable structure in the inner region of a
hydrostatic disk.

Since equation~(\ref{eq59}) is also valid in the case of a freely-falling
inflow, we can combine it with equations~(\ref{eq41}) and (\ref{eq49}) to
conclude that in a {\it freely-falling} disk governed by the Yuan et al.
viscosity, the counterpart of equation~(\ref{eq60}) is given by
\begin{equation}
\beta = {\gamma \over 2} \ \ \ \ \ \ \ 
\phantom{sssssss} {\rm (free-fall)} \ .
\label{eq61}
\end{equation}
We again find that $\beta < 1$ for $4/3 < \gamma < 5/3$, and therefore
the viscosity prescription proposed by Yuan et al. (2000) is unphysical
close to the horizon whether the disk has a free-fall or hydrostatic
structure.

\subsection{Richard \& Zahn}

The prescription proposed by Richard \& Zahn (1999) gives for the viscosity
and the shear stress
\begin{equation}
\nu = - \alpha \, r^3 {d \Omega \over dr} \ , \ \ \ \ \ \ \ 
\Sigma = \alpha \, r^4 \rho \left(d \Omega \over dr\right)^2 \ .
\label{eq62}
\end{equation}
In this scenario, the angular momentum transport is based on a nonlinear
diffusion equation. Equations~(\ref{eq17}) and (\ref{eq62}) can be
combined to obtain
\begin{equation}
B \, (r - \rs)^\beta \asym {\alpha \, r^5 \over v}
\, \left(d\Omega \over dr\right)^2 \ .
\label{eq63}
\end{equation}
Incorporating the asymptotic behavior of $v$ near the event horizon
(equation~\ref{eq41}), and requiring that the powers of $(r-\rs)$ balance
on the two sides gives for $\beta$ the result
\begin{equation}
\beta = {1 \over 2} \ \ \ \ \ \ \ 
{\rm (hydrostatic)} \ .
\label{eq64}
\end{equation}
This result fails to satisfy the existence condition $\beta \ge 1$, and
therefore we must conclude that the Richard-Zahn prescription for the
viscosity is unphysical in the region close to the event horizon if the
disk is in hydrostatic equilibrium.

Equation~(\ref{eq63}) also applies in the case of an accretion disk with
a free-fall inner region. Since the sound speed $a$ does not appear in
this expression, we find that the same result is obtained for $\beta$ in
a freely-falling disk subject to the Richard \& Zahn viscosity, namely
\begin{equation}
\beta = {1 \over 2} \ \ \ \ \ \ \ 
\phantom{s} {\rm (free-fall)} \ .
\label{eq65}
\end{equation}
It therefore follows that the Richard \& Zahn prescription for the variation
of the viscosity $\nu$ yields unphysical structure in the inner region of an
accretion disk, whether the disk is freely falling or in vertical hydrostatic
equilibrium.

\section{CONCLUSIONS}

In this paper we have explored the implications of several alternative
prescriptions for the kinematic viscosity $\nu$ on the structure of an
advection-dominated accretion disk close to the event horizon of a
nonrotating black hole. Our investigation has been motivated primarily
by the results obtained by Becker \& Le (2003), who found that the
original Shakura-Sunyaev (1973) prescription yields acceptable disk
structure, despite its ``diffusive'' nature, which has caused other
authors to propose ``deterministic'' alternatives. Our approach has been
to employ rigorous asymptotic analysis to determine which of the various
prescriptions for $\nu$ considered here satisfy the fundamental
existence conditions for the inflow. In order to make our comparison in
the most model-independent way possible, we have stipulated that the
stress must vanish exactly at the event horizon, which is a fundamental
requirement of general relativity (Weinberg 1972). The specific
scenarios for the viscosity variation we have examined are those
considered by Narayan et al. (1997), Richard \& Zahn (1999), and Yuan et
al. (2000), as well as the Shakura-Sunyaev form. The disk was assumed to
be in either vertical hydrostatic equilibrium or free-fall, and the
pseudo-Newtonian potential was utilized.

We have analyzed the structure of the accretion disk near the event
horizon associated with the Shakura-Sunyaev viscosity prescription in
\S~4 and compared the results with those obtained using the three
alternative formulations of interest here. Interestingly, we find that
none of the three alternatives yields a physically consistent structure
for the accretion disk close to the event horizon. This includes both
the deterministic forms considered by Narayan et al. (1997) and Yuan et
al. (2000), as well as the nonlinear prescription suggested by Richard
\& Zahn (1999). From a physical point of view, these alternative models
fail to satisfy the zero-stress boundary condition at the event horizon
of the black hole. Hence we conclude that the only acceptable form for
the viscosity is the ``diffusive'' prescription originally proposed by
Shakura \& Sunyaev (1973). This conclusion holds both for disks in
hydrostatic equilibrium, as well as for those experiencing free-fall in
the inner region. One implication of our results is that the radial
derivative of the torque vanishes at the horizon, since this is a
property of the Shakura-Sunyaev viscosity formulation (see Becker \& Le
2003).

\subsection{Inner Boundary Condition}

The deterministic viscosity prescriptions considered by Narayan, Kato \&
Honma (1997) and Yuan et al. (2000) result in first-order equations for
the time evolution of $\Omega$, as pointed out in \S~3.2 and \S~3.3. A
first-order differential equation only needs one boundary condition; one
is therefore free to choose either the inner or outer boundary
condition. Since the flow is typically supersonic near the black hole,
Narayan, Kato \& Honma (1997) argue that information cannot propagate
upstream from the region near the event horizon. According to this line
of reasoning, the inner boundary condition is not relevant, and one is
justified in using the outer boundary condition (which corresponds to
conditions far away from the black hole). However, as discussed in
\S~1.1, angular momentum in accretion flows around black holes might
well be transported by magnetic stresses, or by torques arising from
some combination of magnetic fields and particles. In particular, a
number of recent papers dealing with MHD simulations of black hole
accretion flows have emphasized the role of magnetic fields in producing
viscous stresses (Reynolds \& Armitage 2001; Hawley \& Krolik 2001; Agol
\& Krolik 2000; Gammie 1999; Menou 2003; Krolik \& Hawley 2002).

If magnetic fields and/or particles play a significant role in
transmitting torques, then the viscous stress can propagate at the
Alfv\'en speed, or at some other velocity that could well exceed the
local sound speed. Furthermore, several of the papers mentioned above
find that viscous stresses can exist in the supersonic region of the
flow well below the radius of marginal stability, where hydrodynamical
effects are no longer important. It follows that in such situations, the
hydrodynamical sonic point (where the fluid flow velocity equals the
local sound speed) does not represent the fundamental boundary beyond
which mechanical stresses cannot propagate upstream. We therefore argue
that the only essential inner boundary condition for the torque is that
{\it the viscous stress must vanish exactly at the event horizon}, as
mandated by general relativity. The first-order equation for the time
evolution of $\Omega$ associated with the ``deterministic'' viscosity
prescriptions studied by Narayan, Kato \& Honma (1997) and Yuan et al.
(2000) requires one fewer boundary conditions than the second-order
equation for $\Omega$ obtained in the ``diffusive'' scenarios (such as
the one originally proposed by Shakura \& Sunyaev). However, even in
these cases, the zero-stress condition at the event horizon is not
optional, and it must be given preference over less fundamental
conditions applied at a large distance from the black hole.

We suggest that the diffusive nature of the Shakura-Sunyaev form is the
reason it is able to satisfy the zero-stress inner boundary condition.
Hence, rather than representing a drawback, this feature actually
enables the development of a physically consistent flow structure in the
vicinity of the horizon. Furthermore, as demonstrated by Becker \& Le
(2003), the diffusive character of the Shakura-Sunyaev prescription does
not lead to relativistic causality violations close to the event horizon
because signals propagating in that region are advected into the black
hole at the speed of light, as required. In our view, the success of the
original Shakura-Sunyaev formulation reflects the simple fact that the
viscous transport of angular momentum in accretion disks is indeed a
diffusive physical process (Pringle 1981).

\subsection{Kerr Black Holes}

It is interesting to consider how the picture presented here would be
modified if the black hole possessed finite angular momentum, rather
than being nonrotating as we have assumed. While a definitive answer to
this question is beyond the scope of the present paper, we shall
nonetheless make a few general observations and a conjecture. The
location of the sonic point (or points if the disk contains a shock) is
be quite sensitive to the angular momentum of the black hole, as
discussed by Sponholz \& Molteni (1994) and more recently by Barai, Das,
\& Wiita (2004). The values of the various disk structure variables such
as the pressure, density, etc. will therefore depend on both the angular
momentum of the black hole, and also on whether the gas is orbiting in
the prograde or retrograde directions.

In the Schwarzschild metric, the static limit and the event horizon are
both located at radius $\rs = 2GM/c^2$. However, when the black hole
possesses finite angular momentum, frame dragging causes the values of
these two radii to bifurcate (Bardeen, Press, \& Teukolsky 1972). This
complicates efforts to model accretion flows around rotating black holes
using pseudopotentials, but despite this there have been several
attempts to do so (e.g., Sponholz \& Molteni 1994; Mukhopadhyay 2003;
Chakrabarti \& Khanna 1992). For example, the Kerr pseudopotential
analyzed by Chakrabarti \& Khanna (1992) can be written in the form
\begin{equation}
\Phi_{\rm Kerr}(r) = c^2 - {GM \over r - r_*} + {1 \over 2}
\, \omega^2 r^2 +{A_* \rs a_* \ell \over r^3} + {(1-\rs/r)
\ell^2 \over 2 r^2}
\ ,
\label{eq5.1}
\end{equation}
where $a_*$ is the angular momentum per unit mass of the black hole,
$A_*$ is a dimensionless spin-orbit coupling constant, and $\omega$ is
the angular velocity of the metric rotation (see, e.g., Bardeen, Press,
\& Teukolsky 1972). The singularity radius, $r_*$, is set in an ad hoc
manner in order to maximize the agreement with the fully relativistic
calculations. For rapidly-rotating black holes, Chakrabarti \& Khanna
(1992) use $r_* = \pm GM/c^2$, where the plus and minus signs refer to
prograde and retrograde orbits, respectively. In the prograde
case, the potential diverges at the event horizon (Bardeen, Press, \&
Teukolsky 1972), but this is not so in the retrograde case. Hence the
accretion dynamics in these two situations will be quite different.

For disks orbiting in the prograde sense, the Kerr pseudopotential given
by equation~(\ref{eq5.1}) has the same divergent behavior near the
horizon as the nonrotating potential used here if we replace $\rs$ in
equation~(\ref{eq2}) with the actual horizon radius ($GM/c^2$ for a
rapidly rotating hole). When combined with the Newtonian energy equation
employed by Chakrabarti \& Khanna (1992), we therefore conclude that the
radial component of the four-velocity has the same asymptotic behavior
as $v$ in our equation~(\ref{eq9}) if the flow is prograde. Furthermore,
we point out that our equation~(\ref{eq10}) for the specific angular
momentum $\ell$ can be combined with equations~(\ref{eq1}) and
(\ref{eq3}) to obtain
\begin{equation}
\dot M \, \ell + 4 \pi r^2 H \, \Sigma = \dot M \, \ell_0
\ ,
\label{eq5.2}
\end{equation}
which is identical to equation~(52) in the general relativistic
treatment of Gammie \& Popham (1998), aside from a sign difference in
the definition of the shear stress. Based on these conceptual
similarities, we expect that the asymptotic structure of a prograde
accretion disk (close to the event horizon) in the Kerr case is likely
to be similar to that derived here for nonrotating black holes. If this
turns out to be true, then the conclusions we have reached regarding the
``existence conditions'' for the various vicosity prescriptions will
also hold for prograde disks around Kerr black holes. This conjecture
clearly needs to be checked in the future using a detailed quantitative
calculation. On the other hand, the pseudopotential in the retrograde
case has a completely different behavior near the horizon (i.e., the
potential is not divergent), and therefore one cannot draw a direct
analogy with the work we have presented here for nonrotating holes. We
therefore defer further discussion of this case to a subsequent paper.

\subsection{Discussion}

Our results have direct relevance for the computational modeling of the
structure of accretion disks around black holes, since the prescriptions
for the viscosity that fail to satisfy the existence conditions
developed here cannot be used as the basis for self-consistent accretion
models close to the event horizon. Because our approach is based on a
careful consideration of the fundamental boundary conditions for the
accretion flow near the event horizon, our results provide insight into
how these conditions constrain the global structure of the accretion
flow. This is particularly important when one examines the relation
between inflow (accretion) and the powerful outflows (winds and jets)
commonly observed to emanate from radio-loud systems containing black
holes. These outflows may be powered by particle acceleration occurring
at a standing, centrifugally-supported shock in the underlying accretion
disk (Le \& Becker 2004; Yuan et al. 2002; Chakrabarti 1989; Abramowicz
\& Chakrabarti 1990; Chakrabarti \& Das 2004; Lu, Gu, \& Yuan 1999;
Chakrabarti 1990). The location of the shock is determined by the
conservation equations and the shock jump conditions, along with the
boundary conditions. Hence the location of the shock (and therefore its
Mach number, compression ratio, etc.) depends on the behavior of the
accretion flow close to the event horizon, since this determines the
inner boundary conditions. It follows that a firm understanding of this
behavior is essential in order to develop self-consistent global models
for the disk/shock/outflow system.

In this work we have utilized the pseudo-Newtonian potential in lieu of
a full treatment of general relativity, in contrast to the work of Das
(2004) and Barai et al. (2004). However, many studies in the literature
have confirmed that this potential provides remarkably good agreement
with the predictions of full general relativity, even very close to the
event horizon (e.g., Becker \& Le 2003; Paczy\'nski \& Wiita 1980). In
particular, the dynamics of free particles close to the horizon in the
pseudo-Newtonian potential agrees exactly with the relativistic results.
Hence we are confident that the conditions derived here are fully
applicable in the relativistic case. By utilizing the most conservative
possible boundary condition for the stress (that it vanish at the
horizon), we have obtained general results that facilitate the critical
evaluation of the various forms for the viscosity that have been
proposed in the literature. Our focus in this paper has been on the
asymptotic behavior close to the event horizon of ``perfect''
advection-dominated disks, which lose no matter or energy. However, we
argue that our conclusions will also apply to disks that lose matter and
energy (whether ADAF or not), provided the losses do not occur very
close to the horizon.

The authors are grateful to the anonymous referee for several
stimulating suggestions that led to significant improvements in the
presentation and discussion. PAB would also like to acknowledge
generous support provided by the Naval Research Laboratory during a
portion of this research.


\label{lastpage}

\end{document}